\documentclass[10pt,twocolumn,english,aps,prd,nofootinbib,preprintnumbers]{revtex4}
\usepackage{amssymb,amsmath,amsfonts,amsbsy,epsfig,graphicx}
\usepackage[dvipsnames]{xcolor}
\definecolor{lgray}{gray}{0.35}
\usepackage[colorlinks=true,urlcolor=Gray,linkcolor=lgray,citecolor=Gray, pdfpagelabels=true,hypertexnames=true,plainpages=false,naturalnames=false]{hyperref}

\newcommand{\be}{\begin{equation}}
\newcommand{\ee}{\end{equation}}
\newcommand{\bea}{\begin{eqnarray}}
\newcommand{\eea}{\end{eqnarray}}
\newcommand{\nn}{\nonumber}
\def\lcdm{$\Lambda$CDM }
\def\C{{\mathcal C}}

\def\op{{\Omega_{\rm PBH}}}

\def\oPBH{\Omega_{\rm PBH}}
\def\mPBH{M_{\rm PBH}}
\def\omx{\Omega_{\rm x}}
\begin{document}

\title{Evaporating primordial black holes as varying dark energy}

\date{\today}

\author{Savvas Nesseris}\email{savvas.nesseris@csic.es}
\affiliation{Instituto de F\'isica Te\'orica UAM-CSIC, Universidad Auton\'oma de Madrid,
Cantoblanco, 28049 Madrid, Spain}

\author{Domenico Sapone}\email{domenico.sapone@uchile.cl}
\affiliation{Grupo de Cosmolog\'ia y Astrof\'isica Te\'orica, Departamento de F\'{i}sica, FCFM, \mbox{Universidad de Chile}, Blanco Encalada 2008, Santiago, Chile.}

\author{Spyros Sypsas}\email{s.sypsas@gmail.com}
\affiliation{Department of Physics, Faculty of Science, Chulalongkorn University, Phayathai Rd., Bangkok 10330, Thailand}

\begin{abstract}
If light enough primordial black holes (PBH) account for dark matter, then its density decreases with time as they lose mass via Hawking radiation. We show that this time-dependence of the matter density can be formulated as an equivalent $w(z)$ dark energy model and we study its implications on the expansion history. Using our approach and comparing with the latest cosmological data, including the supernovae type Ia, Baryon Acoustic Oscillations, Cosmic Microwave Background and the Hubble expansion H(z) data, we place observational constraints on the PBH model. We find that it is statistically consistent with \lcdm according to the AIC statistical tool. Furthermore, we entertain the idea of having a population of ultra-light PBHs, decaying around neutrino decoupling, on top of the dark matter fluid and show how this offers a natural dark matter-radiation coupling altering the expansion history of the Universe and alleviating the $H_0$ tension.
\end{abstract}

\maketitle

\section{Introduction}\label{sec:intro}
The recent observation of gravitational wave emission from black hole inspirals~\cite{Abbott:2016blz} has revitalized the tantalizing idea that a fraction of our Universe's dark matter (DM) budget could consist of primordial black holes (PBH)~\cite{Hawking:1971ei}. The unexpectedly large black hole merger rate~\cite{Abbott:2016nhf} inferred from LIGO observations overlaps with various estimations derived from PBH dark matter models~\cite{Bird:2016dcv}. This view is reinforced by the fact that the progenitor black holes responsible for the emission of gravitational waves seem to be spinless~\cite{Abbott:2016blz}, a property which is unlikely to be found in black holes of astrophysical origin, while natural for PBHs~\cite{Garcia-Bellido:2017fdg,Clesse:2017bsw,Mirbabayi:2019uph,DeLuca:2019buf}.

The PBH idea is reshaping our understanding of dark matter as the formation of PBHs shortly after the Big-Bang, during the radiation era, requires primordial curvature fluctuations with amplitudes large enough to induce the gravitational collapse of matter into black holes upon re-entering the horizon. A well-known mechanism leading to such an amplification is a period of ultra slow-roll inflation~\cite{Garcia-Bellido:2017mdw,Germani:2017bcs}, due to an inflection point as in, for example, Higgs inflation~\cite{Ezquiaga:2017fvi}. Furthermore, PBHs might have been produced in abundance due to various other mechanisms. Examples include: phase transitions, e.g., bubble collisions~\cite{Crawford:1982yz,Hawking:1982ga,Kodama:1982sf,Moss:1994iq,Konoplich:1999qq}, collapse of topological defects~\cite{Hogan:1984zb,Hawking:1987bn,Polnarev:1988dh,Garriga:1993gj,Rubin:2000dq,Khlopov:2000js,Rubin:2001yw} or other solitonic objects~\cite{Cotner:2016cvr}; a period of slow reheating ---i.e. an early matter dominated era~\cite{Khlopov:1985jw,Carr:1994ar}; resonance effects~\cite{Cai:2018tuh}; instabilities of fifth force mediators~\cite{Amendola:2017xhl}; strongly coupled cosmologies~\cite{Bonometto:2018dmx}. Finally, being sensitive to the tail distribution of primordial fluctuations~\cite{Young:2013oia,Atal:2018neu,Atal:2019cdz,Panagopoulos:2019ail}, the PBH formation is intertwined with non-Gaussian initial conditions relating the dark matter density to the dynamics of the primordial Universe.

On the other hand, a tension has emerged between long~\cite{Aghanim:2018eyx} and short (local)~\cite{Riess:2018byc} distance measurements of the Hubble constant, which has persisted up to now at a considerably high confidence level of 4.4$\sigma$~\cite{Riess:2019cxk}, a possibility that, if confirmed, could open the way for new physics at cosmological, or possibly, even microphysical scales.

It is well known~\cite{Bernal:2016gxb} that a way to reconcile local and cosmological measurements of $H_0$ is to consider a coupling between dark matter and (dark) radiation~\cite{DiValentino:2015ola,Kumar:2017dnp,DiValentino:2017oaw,DiValentino:2016hlg,DEramo:2018vss,Kreisch:2019yzn,DiValentino:2017iww,DiValentino:2017zyq,Yang:2018euj,Yang:2018uae,Yang:2018qmz,Mortsell:2018mfj,Guo:2018ans,Poulin:2018cxd,Pandey:2019plg,Vattis:2019efj,Yang:2019uzo,Kumar:2019wfs,DiValentino:2019exe} so that the former decays/annihilates to the latter. This process lowers the redshift of matter/radiation equality, thus amplifying the age of the Universe.

In this article, we offer a realization of this scenario due the quantum effect of Hawking evaporation of primordial black holes. In such a case, the radiation, however, is not dark since it is composed of relativistic Standard Model particles and as such it is subjected to constraints. The photons and charged leptons emitted by sufficiently low mass BHs are constrained from $\gamma$- and cosmic-ray observations~\cite{Carr:2016hva,Boudaud:2018hqb}, while neutrino emission has been studied in~\cite{Bugaev:2008gw}. The lowest possible mass for which DM can be entirely due to PBHs is the so-called \emph{asteroid} window $\mPBH\simeq 2\times10^{16}$ g~\cite{Carr:2017jsz}, however, the concensus on the constraints is not fully settled~\cite{Carr:2019yxo}.

In this paper we show that for a monochromatic PBH population around this mass range, the emitted particles behave as a dark energy fluid with a time dependent equation of state $w_{\textrm{DE}}(z)$. We take into account the whole expansion history from recombination to very low redshifts and show that, in this case, $w_{\textrm{DE}}(z)$ mildly crosses the phantom line $w_{\textrm{DE}}(z)=-1$. The effect is too small to have any significant impact on, e.g., the value of $H_0$. However, in Sec.~E, we entertain the idea of having a very small fraction ($f_{\rm PBH}\sim10^{-7}$) of very light primordial black holes ($M_{\rm PBH}\sim10^9$ g), compatible with constraints in this range~\cite{Carr:2009jm}. Such a population would have completely decayed before nucleosynthesis and as such it cannot serve as a DM candidate; it can, however, produce just enough radiation to raise the Hubble constant, thus reducing the tension.

\section{Hubble parameter and horizon at drag epoch}
In this Section we will examine the effect of the black hole mass loss on the energy budget of the Universe. Page in~\cite{Page:1976df}, showed how a BH emits mass as a function of cosmic time, which when written in terms of the scale factor reads
\be
M_{\rm PBH}'(a) = -\frac{\C}{a\,H(a)\,M_{\rm PBH}(a)^2}\,,
\label{eq:eq:mprime}
\ee
where prime denotes the derivative with respect to the scale factor, $H(a)$ is the Hubble parameter and $\C$ is a numerical constant\footnote{${\C}\equiv 3 F(M) \frac{\hbar c^4}{G^2}$, where $F(M)$ is a numerical coefficient that takes into account the distribution of power in different species~\cite{MacGibbon:1991tj}. For the mass range $5\times10^{14} \lesssim M/{\rm g}\lesssim 10^{16}$, where emission occurs via neutrinos, electrons/positrons and photons it has the value $F(M)=4.427 \times 10^{-4}$~\cite{Page:1983ug}.} with dimension $[M^3/t]$. Let us assume that some fraction $f_{\rm PBH}$ of DM is in the form of primordial black holes, so we set $\Omega_{\rm PBH} = f_{\rm PBH} \Omega_{c}$ at some initial time $a=a_{\rm in}$. Hence, the PBH density will read
\be\label{eq:eq:Om}
\oPBH(a) = f_{\rm PBH}\frac{\Omega_{c,0}}{a^3}\frac{M_{\rm PBH}(a)}{M_{\rm in}},
\ee
where $M_{\rm in}$ is the initial mass of the black hole population.

Then we need to consider the loss in $\Omega_{\rm PBH}$ and the simultaneous gain in $\Omega_{\rm x}$ which will be our ``dark'' component.
Clearly, an energy loss in the PBH section will imply a gain in the new dark radiation component, so this will translate to a system of coupled fluids ---see Ref.~\cite{Barrow:1991dn} for a similar approach studying BH evaporation. Following the approach in~\cite{Majerotto:2004ji}, the system of coupled matter  and dark radiation fluid will be given by
\bea
\rho'_c+\frac{3}{a}\rho_c &=&Q(a),\nonumber \\
\rho'_{\rm x}+\frac{3}{a}(1+w_{\rm x})\rho_{\rm x} &=& -Q(a),
\label{eq:system-coupled-t}
\eea
where the positive sign in the matter section refers to a loss in energy. The function $Q$ appearing in the above equations is a coupling term that remains to be specified. Note that, the value of the equation of state parameter is for the moment left unspecified.

The dynamics of the primordial black hole density follows directly from inserting Eq.~\eqref{eq:eq:Om} into Eq.~\eqref{eq:eq:mprime}, to find
\be
\oPBH'(a) +\frac{3}{a}\oPBH(a) =  -\frac{\C/M^3_{\rm in}}{a^{10}\,H}\frac{(f_{\rm PBH}\Omega_{c,0})^3}{\oPBH(a)^2}.
\ee
The dark radiation instead evolves as
\be
\Omega_{\rm x}'(a) +\frac{3(1+w_{\rm x})}{a}\Omega_{\rm x}(a) =  \frac{\C/M^3_{\rm in}}{a^{10}\,H}\frac{(f_{\rm PBH}\Omega_{c,0})^3}{\oPBH(a)^2}\,.
\ee
This system of equations can be further simplified in order to drop the dependence $a^{-10}$, which might lead to instabilities at early times; by considering the normalized energy densities
\bea
\tilde{\Omega}_{\rm PBH}(a) &=& \oPBH(a)\,a^{3},\\
\tilde{\Omega}_{\rm x}(a) &=& \Omega_{\rm x}(a)\,a^{3(1+w_{\rm x})}\,,
\eea
the final system reads
\bea
\tilde{\Omega}_{\rm PBH}'(a) &=& -\frac{\alpha}{a\,H}\frac{(f_{\rm PBH}\Omega_{c,0})^3}{\tilde{\Omega}_{\rm PBH}(a)^2}\label{eq:oPBH_tilde},\\
\tilde{\Omega}_{\rm x}'(a) &=& \frac{\alpha}{a^{1-3w_{\rm x}}\,H}\frac{(f_{\rm PBH}\Omega_{c,0})^3}{\tilde{\Omega}_{\rm PBH}(a)^2}\label{eq:ox_tilde}\,,
\eea
where we have set $\alpha = \C/M^3_{\rm in}$.

We demand that at early times the Universe behave as in the $\Lambda$CDM cosmology, hence, the initial conditions for each species will be
$\tilde{\Omega}_{\rm PBH}(a_{\rm in}) = f_{\rm PBH} \Omega_{c,0}$ and $\tilde{\Omega}_{\rm x}(a_{\rm in}) = 0$.
The problem here is the presence of the Hubble parameter in Eqs.~\eqref{eq:oPBH_tilde} and \eqref{eq:ox_tilde}, which is only defined implicitly at this time. To this end, we impose the Hubble parameter to be exactly the $\Lambda$CDM one, see the Appendix for more details.
The final Hubble parameter will be
\bea
H(a)^2&=&H_0^2\left[\Omega_{r,0}a^{-4}+\Omega_{b_0}a^{-3}+(1-f_{\rm PBH})\Omega_{c_0}a^{-3}+\right.\nonumber \\
&  & \left.  \op(a)+\Omega_{\rm x}(a)+ \Omega_{\Lambda_0} \right]
\label{eq:Hubble-op}
\eea
where
\bea
\Omega_{r,0}&=&\Omega_{\gamma,0}\left(1+N_{\rm eff}\cdot \frac78\cdot\left(\frac{4}{11}\right)^{\frac43}\right) \nonumber \\
\Omega_{\Lambda_0} &=& 1-\Omega_{b_0}-(1-f_{\rm PBH})\Omega_{c_0}-\nonumber \\
&-&\Omega_{r_0}-\op(a=1)-\Omega_{\rm x}(a=1)\,.\nonumber
\eea

\subsection{Effective equation of state parameter}

The PBH radiative mechanism can be also interpreted as an effective dark energy fluid. The emitted particles, even though relativistic, do not behave as radiation, since $\omx$ does not scale as $a^{-4}$ due to Eq.~\eqref{eq:eq:mprime}. For light PBHs, by solving the system of coupled fluids numerically, it can be shown that at late times the dark radiation component becomes comparable to the photon density having a subdominant effect on the dynamics which however excibits a phantom behavior.

The Hubble parameter in a flat, dark energy dominated Universe, is given by
\be
\frac{H^2(a)}{H_0^2} = \Omega_{r,0}a^{-4}+\Omega_{m_0}a^{-3}+\Omega_{\rm DE,0}e^{-3\int_1^{a}\frac{1+w(y)}{y}{\rm d}y},\nonumber
\ee
where $\Omega_{m_0} = \Omega_{b_0}+\Omega_{c_0}$ and $\Omega_{\rm DE,0} = 1-\Omega_{\rm r,0}-\Omega_{\rm m,0}$.
Then the effective equation of state parameter can be expressed in terms of the Hubble parameter as
\be
w_{\rm eff}(a) = \frac{2a H(a)H'(a)/3+1/3 H_0^2\Omega_{r,0}a^{-4}+H^2(a)}{ H_0^2\Omega_{r,0}a^{-4}+ H_0^2\Omega_{m,0}a^{-3}-H^2(a)}\,.\nonumber
\ee
Using Eq.~\eqref{eq:Hubble-op}, we find
\be
w_{\rm eff}(a)= \frac{-w_{\rm x}\Omega_{\rm x}(a)+\Omega_{\Lambda,0}}{f_{\rm PBH}\Omega_{c,0}a^{-3}-\Omega_{\rm PBH}(a)-\omx(a)-\Omega_{\Lambda,0}}\,.
\label{eq:w_eff_PBH}
\ee
As previously stated, we demand the Universe to be $\Lambda$CDM at early times, hence $\Omega_{\rm x}\rightarrow 0$ and $\Omega_{\rm PBH}\rightarrow f_{\rm PBH}\Omega_{c,0}a^{-3}_{\rm in}$: this implies $w_{\rm eff}\rightarrow -1$. At late times, $\Omega_{\rm x}>0$, as black holes are producing more relativistic matter by losing their masses, $\Omega_{\rm PBH}<f_{\rm PBH}\Omega_{c,0}$. If we rewrite Eq.~\eqref{eq:w_eff_PBH} as
\be
w_{\rm eff}(1)=-1 +\frac{f_{\rm PBH}\Omega_{c,0}-\Omega_{\rm PBH}(1)-(w_{\rm x}+1)\Omega_{\rm x}(1)}{f_{\rm PBH}\Omega_{c,0}-\Omega_{\rm PBH}(1)-\omx(1)-\Omega_{\Lambda,0}} \nonumber
\ee
we realize that the denominator is always negative because $\Omega_{\Lambda,0}$ is the largest component at late times; both terms in the numerator are positive, however the ``dark'' radiation is subdominant with respect to the PBH energy loss due to the $\approx a^{-4}$ scaling of the solution in the radiation sector. This guarantees the fraction to be always negative, implying an effective equation of state parameter always less than $-1$.

However, the effective phantom behavior is very mild for the sensible value of the PBH masses; for instance, if we assume a mass of $10^{16}$g, the effective equation of state parameter differs from $-1$ of about $0.001\%$. Only for very light primordial black holes, i.e. $M_{\rm PBH} \sim 10^{15}$g, $w_{\rm eff}(1)= -1.01$. The abundance of such black holes, however, is severely constrained.

\section{Cosmological constraints}
\subsection{Data}
In this section, we will now present the parameter constraints from fitting our model to the latest cosmological data such as the supernovae type Ia (SnIa), Baryon Acoustic Oscillations (BAO), Cosmic Microwave Background (CMB) and the Hubble expansion H(z) data. Specifically, we utilize the Pantheon Type Ia Supernovae (SnIa) compilation of Ref.~\cite{Scolnic:2017caz}, the BAO measurements from 6dFGS~\cite{Beutler:2011hx}, SDDS~\cite{Anderson:2013zyy}, BOSS CMASS~\cite{Xu:2012hg}, WiggleZ~\cite{Blake:2012pj}, MGS~\cite{Ross:2014qpa}, BOSS DR12~\cite{Gil-Marin:2015nqa} and DES Y1~\cite{Abbott:2017wcz}. Finally, we also use the CMB shift parameters $(R, l_a)$ that are based on the \textit{Planck 2018} release~\cite{Aghanim:2018eyx}.

Moreover, we also incorporate in our analysis the direct measurements of the Hubble expansion $H(z)$ data. These can be derived in two ways: via the differential age method or by the clustering of galaxies and quasars. The latter provides direct measurements of the Hubble parameter by measuring the BAO peak in the radial direction from the clustering of galaxies or quasars~\cite{Gaztanaga:2008xz}. On the other hand, the former method determines the Hubble parameter via the redshift drift of distant objects over significant time periods, usually a decade or longer. This is possible as in GR the Hubble parameter can be expressed via of the rate of change of the redshift $H(z)=-\frac{1}{1+z}\frac{{\rm d}z}{{\rm d}t}$~\cite{Jimenez:2001gg}. Both of these methods then provide us with a compilation of 36 Hubble parameter $H(z)$ data points, which for completeness we present in Table~\ref{tab:Hzdata}, along with their references.

\subsection{CMB likelihood}
Here we provide some more details about our CMB shift parameters likelihood, where we mostly follow Ref.~\cite{Zhai:2018vmm}. Since we are interested in constraining the extra relativistic degrees of freedom, we rederive the shift parameters with $N_{\textrm{eff}}$ as a free parameter. The shift parameters are given by
\bea
R&\equiv& \sqrt{\Omega_{m0} H_0^2}~r(z_*)/c,\\
l_a&\equiv& \pi~r(z_*)/r_s(z_*),
\eea
where the comoving sound horizon is defined as
\be
r_s(z)=\int_0^a\frac{{\rm d}a'}{\sqrt{3(1+R_b a')a'^4 E(a')^2}}\,.\nonumber
\ee
Here, $E(a)=H(a)/H_0$ is the dimensionless Hubble parameter, $R_b=3\rho_b/4 \rho_r\cdot a^{-1}$ is the baryon-photon ratio and the comoving distance is
\be
r(z)=\frac{c}{H_0}\int_0^z\frac{{\rm d}z'}{E(z)}\,.\nonumber
\ee
Then, we can use these definitions to obtain the new values for the set of shift parameters by using the  MCMC Planck 18 chain\footnote{The chain used is: \\``base\_nnu\_mnu\_plikHM\_TTTEEE\_lowl\_lowE\_post\_lensing".}, see~\cite{Aghanim:2018eyx} for details. To do this, we calculate the parameter vector $\textbf{v}_{\textrm{CMB}}=(R,l_a,\Omega_{b0}h^2, N_{\textrm{eff}})$ for all points in the chain, which gives the mean values
\be
\textbf{v}_{\textrm{CMB},data}=
\left(
  \begin{array}{c}
    1.75478 \\
    302.347 \\
    0.0222369 \\
    2.92029 \\
  \end{array}
\right)\,. \label{eq:CMBshift}
\ee
The effective number of relativistic neutrinos $N_{\rm eff}$ should be modified according to the black hole emission of relativistic species, i.e.  $N_{\textrm{eff}}=N_{\textrm{eff,SM}}+\Delta N_{\textrm{eff}}$. The total $N_{\rm eff}$ will be the sum of both contributions from the
Standard Model particle, denoted by ``SM" and the PBH one, given by $\Delta N_{\textrm{eff}}$, which is directly connected to the
$\Omega_{\rm x}(a)$. However, our choice is more conservative and we decide to leave $N_{\rm eff}$ as a free parameter.
Finally, the covariance matrix of these parameters is given by:
\bea
C_{ij}&=&10^{-8}\times \nn\\
&&\left(\begin{array}{cccc}
    7976.97 & 298535.0 & -137.736 & -75126.6 \\
    298535.0 & 2.11245\cdot10^7 & -1615.98 & 881421.0 \\
    -137.736 & -1615.98 & 5.24732 & 3257.56 \\
    -75126.6 & 881421.0 & 3257.56 & 3.7128\cdot10^6 \\
  \end{array}\right). \nn
\eea
Combining all of the above, the CMB contribution to the $\chi^2$ becomes
\be
\chi^{2}_{\rm CMB}=\delta \textbf{v}~C^{-1}_{CMB,ij}~\delta \textbf{v},
\ee
where the parameter vectors are given by
\be
\delta \textbf{v}=\textbf{v}_{\textrm{CMB}}-\textbf{v}_{\textrm{CMB},data}.
\ee

\begin{table}[!t]
\caption{The $H(z)$ data used in the current analysis (in units of $\textrm{km}~\textrm{s}^{-1} \textrm{Mpc}^{-1}$). This compilation is partly based on those of Refs.~\cite{Moresco:2016mzx} and~\cite{Guo:2015gpa}.\label{tab:Hzdata}}
\small
\centering
\begin{tabular}{cccccccc}
\\
\hline\hline
$z$  & $H(z)$ & $\sigma_{H}$ & Ref.   & $z$  & $H(z)$ & $\sigma_{H}$ & Ref.   \\
\hline
$0.07$    & $69.0$   & $19.6$  & \cite{Zhang:2012mp} & $0.48$    & $97.0$   & $62.0$  & \cite{STERN:2009EP}   \\		
$0.09$    & $69.0$   & $12.0$  & \cite{STERN:2009EP} & $0.57$    & $96.8$   & $3.4$   & \cite{Anderson:2013zyy}  \\	
$0.12$    & $68.6$   & $26.2$  & \cite{Zhang:2012mp} & $0.593$   & $104.0$  & $13.0$  & \cite{MORESCO:2012JH}  \\	
$0.17$    & $83.0$   & $8.0$   & \cite{STERN:2009EP}  & $0.60$    & $87.9$   & $6.1$   & \cite{Blake:2012pj}   \\		
$0.179$   & $75.0$   & $4.0$   & \cite{MORESCO:2012JH} & $0.68$    & $92.0$   & $8.0$   & \cite{MORESCO:2012JH}    \\		
$0.199$   & $75.0$   & $5.0$   & \cite{MORESCO:2012JH} & $0.73$    & $97.3$   & $7.0$   & \cite{Blake:2012pj}   \\	
$0.2$     & $72.9$   & $29.6$  & \cite{Zhang:2012mp}   & $0.781$   & $105.0$  & $12.0$  & \cite{MORESCO:2012JH} \\	
$0.27$    & $77.0$   & $14.0$  & \cite{STERN:2009EP}  & $0.875$   & $125.0$  & $17.0$  & \cite{MORESCO:2012JH} \\	
$0.28$    & $88.8$   & $36.6$  & \cite{Zhang:2012mp}  &	$0.88$    & $90.0$   & $40.0$  & \cite{STERN:2009EP}   \\	
$0.35$    & $82.7$   & $8.4$   & \cite{Chuang:2012qt}  & $0.9$     & $117.0$  & $23.0$  & \cite{STERN:2009EP}   \\	
$0.352$   & $83.0$   & $14.0$  & \cite{MORESCO:2012JH}	& $1.037$   & $154.0$  & $20.0$  & \cite{MORESCO:2012JH} \\
$0.3802$  & $83.0$   & $13.5$  & \cite{Moresco:2016mzx}	& $1.3$     & $168.0$  & $17.0$  & \cite{STERN:2009EP}   \\
$0.4$     & $95.0$   & $17.0$  & \cite{STERN:2009EP}   & $1.363$   & $160.0$  & $33.6$  & \cite{Moresco:2015cya}  \\			
$0.4004$  & $77.0$   & $10.2$  & \cite{Moresco:2016mzx} & $1.43$    & $177.0$  & $18.0$  & \cite{STERN:2009EP}   \\	
$0.4247$  & $87.1$   & $11.2$  & \cite{Moresco:2016mzx} & $1.53$    & $140.0$  & $14.0$  & \cite{STERN:2009EP}  \\	
$0.44$    & $82.6$   & $7.8$   & \cite{Blake:2012pj}   	& $1.75$    & $202.0$  & $40.0$  & \cite{STERN:2009EP}  \\		
$0.44497$ & $92.8$   & $12.9$  & \cite{Moresco:2016mzx}	& $1.965$   & $186.5$  & $50.4$  & \cite{Moresco:2015cya}  \\
$0.4783$  & $80.9$   & $9.0$   & \cite{Moresco:2016mzx} & $2.34$    & $222.0$  & $7.0$   & \cite{Delubac:2014aqe}   \\	
\hline\hline
\end{tabular}
\end{table}

\begin{table*}[!t]
\caption{The best-fit parameters for the $\Lambda$CDM and the PBH models respectively. \label{tab:bestfits}}
\begin{centering}
\begin{tabular}{ccccccc}
\hline
 & $h$ & $N_{\rm eff}$ & $\Omega_{b,0}$ & $\Omega_{c,0} $ & $\log_{10} 	\alpha$& $\log_{10}f_{\rm PBH}$  \\
\hline
\multicolumn{7}{c}{$\Lambda$CDM }\\
\hline
CMB & $0.6536 \pm 0.0189$ &  $2.9202 \pm 0.2009$ & $0.0520 \pm 0.0026$ & $0.2801 \pm 0.0157$ & $-$ & $-$ \\
\hline
CMB+loc & $0.6697 \pm 0.0087$ & $2.9749 \pm 0.1400$ & $0.0499 \pm 0.0011$  & $0.2642 \pm 0.0058$ & $-$& $-$ \\
\hline
\multicolumn{7}{c}{PBH fixing $\log_{10}\alpha$ and $\log_{10}f_{\rm PBH}$}\\
\hline
CMB & $0.6535 \pm 0.0173$ & $2.9117 \pm 0.1706$ & $0.0521 \pm 0.0024$ & $0.2796 \pm 0.0156$ & $-$ & $-$ \\
\hline
CMB+loc & $0.6703 \pm 0.0086$ & $2.9771 \pm 0.1433$ & $0.0499 \pm 0.0011$ & $0.2637 \pm 0.0057$ & $-$ & $-$ \\
\hline
\multicolumn{7}{c}{PBH}\\
\hline
CMB & $0.6537 \pm 0.0192$ & $2.9221 \pm 0.1452$ & $0.0520 \pm 0.0022$ & $0.2802 \pm 0.0139$ & $-10.1349 \pm 5.6832$ & $0.0571 \pm 5.8963$ \\
\hline
CMB+loc & $0.6709 \pm 0.0176$ & $2.9980 \pm 0.1445$ & $0.0498 \pm 0.0018$ & $0.2640 \pm 0.0104$ & $-0.1933 \pm 5.8146$ & $-12.1575 \pm 5.9284$ \\
\hline
\end{tabular}
\par
\end{centering}
\end{table*}

\begin{table}[!t]
\caption{The $\chi^2$ and AIC parameters for the $\Lambda$CDM and the PBH models respectively. \label{tab:chi2AIC}}
\begin{centering}
\begin{tabular}{cccc}
Model & $\chi^2$ & AIC & $\Delta$AIC \\\hline
\hline
\multicolumn{4}{c}{CMB only}\\
\hline
$\Lambda$CDM & $0.02 $ & $ -5.9322 $ &$ 0 $ \\
\hline
PBH          & $ 0.01$  & $ -5.9322 $ &$ 0 $ \\
\hline
PBH  full     & $ 0.003$  & $ -4.6504$ &$ 1.2818 $ \\
\hline
\hline
\multicolumn{4}{c}{CMB+$H(z)$+SNIa+BAO}\\
\hline
$\Lambda$CDM & $1079.60 $ & $ -24.2354 $ &$ 1.9437 $ \\\hline
PBH          & $ 1079.59$  & $ -22.2917 $ &$ 0 $ \\\hline
PBH full      & $ 1079.63$  & $ -29.9687 $ &$ 7.677 $ \\\hline

\end{tabular}
\par
\end{centering}
\end{table}

\begin{figure*}[!t]
\centering
\includegraphics[width = 0.8\textwidth]{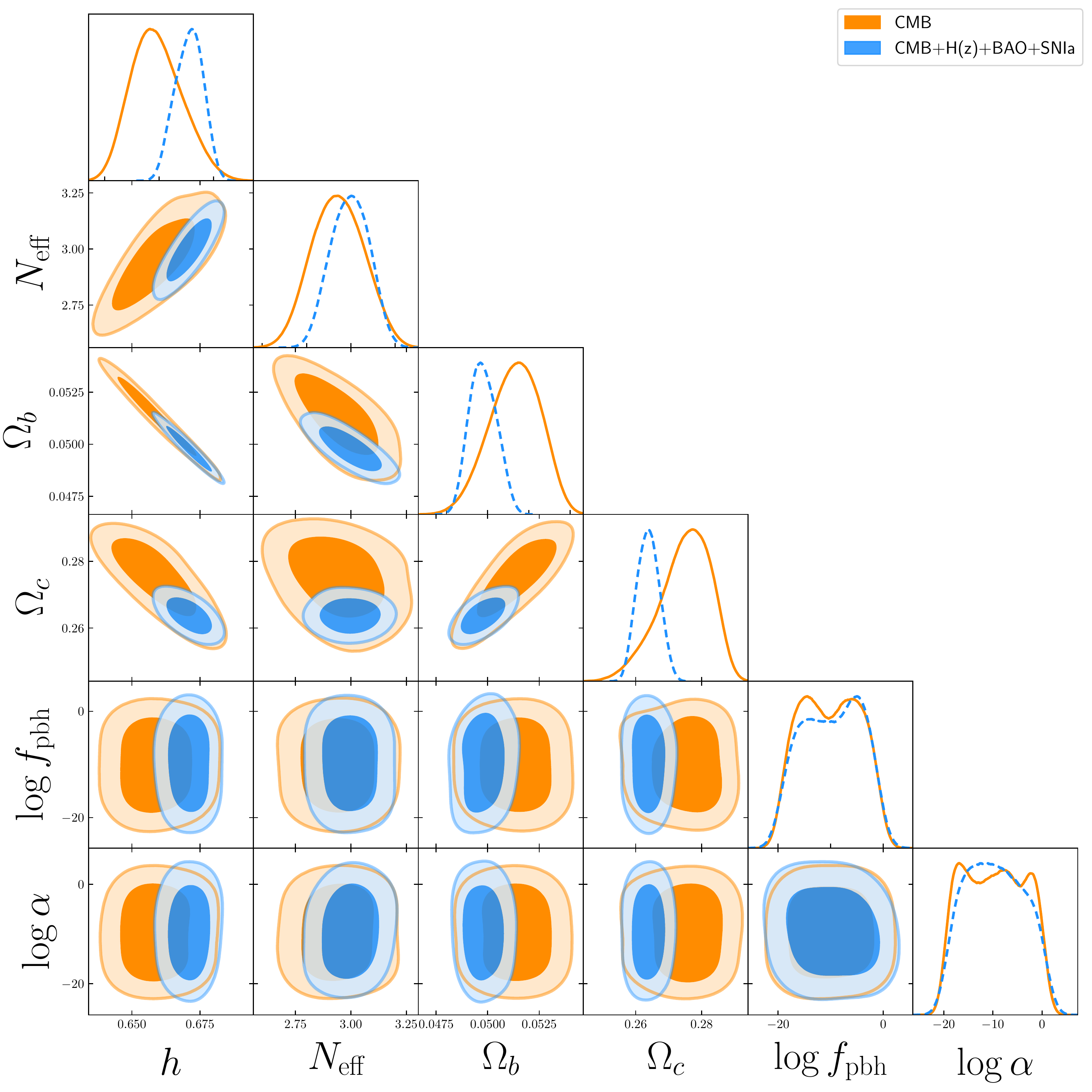}
\caption{The 68.3$\%$ and 95.4$\%$  confidence contours and the 1D marginalized likelihoods or various parameter combinations for the PBH model.
\label{fig:MCMCPBH-full}}
\end{figure*}

\begin{figure*}[!t]
\centering
\includegraphics[width = 0.8\textwidth]{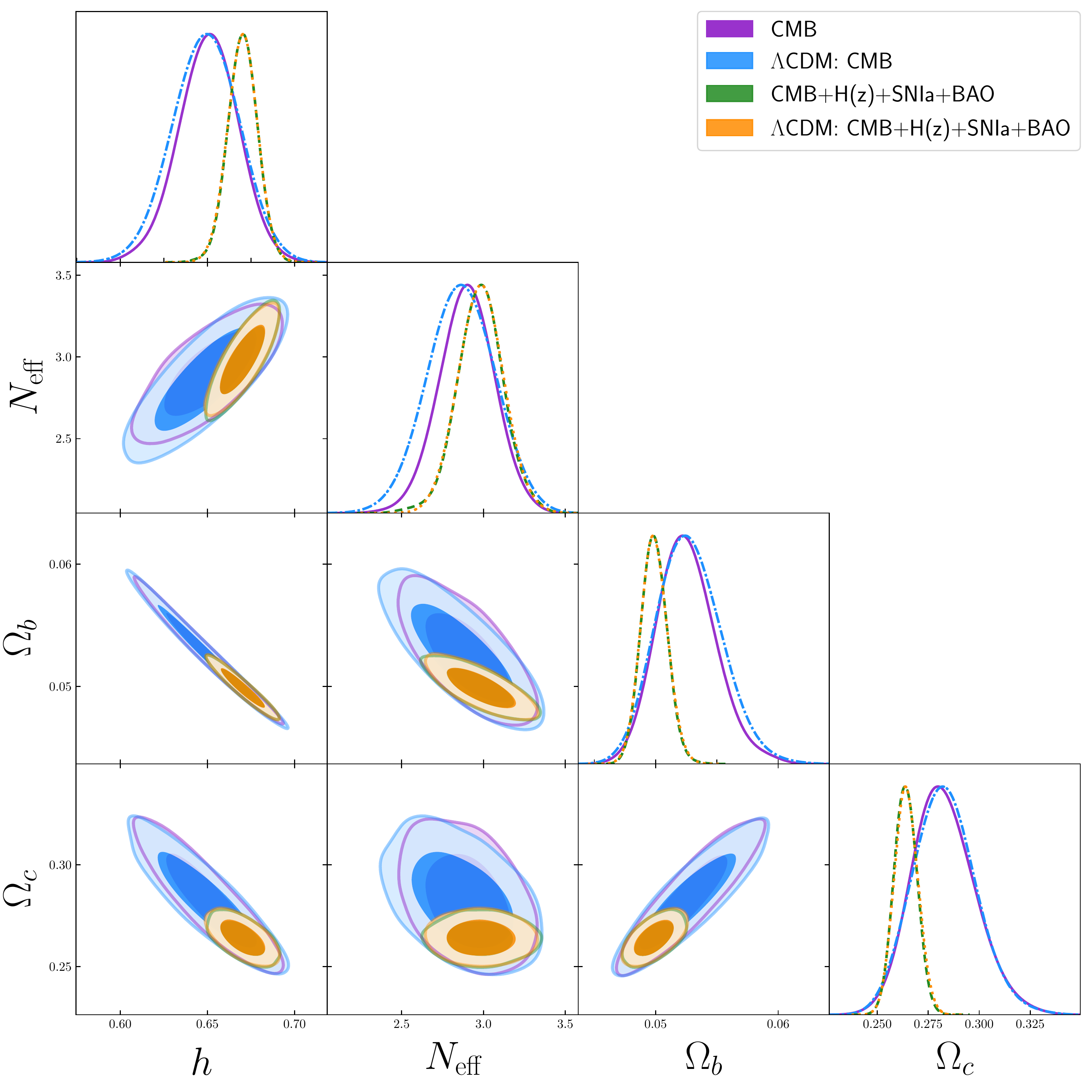}
\caption{The 68.3$\%$ and 95.4$\%$  confidence contours and the 1D marginalized likelihoods or various parameter combinations for the PBH model.
\label{fig:MCMCPBH}}
\end{figure*}

\subsection{Methodology}
In order to use the aforementioned data we first need to estimate the background expansion history of the Universe by calculating the Hubble parameter. This can be achieved by solving Eq.~\eqref{eq:Hubble-op} together with Eqs.~\eqref{eq:oPBH_tilde} and \eqref{eq:ox_tilde}.

From Eq.~\eqref{eq:Hubble-op} we can easily calculate the necessary cosmological distances required by the data using the usual FRW definitions. Then, our total likelihood function $L_{\rm tot}$ can be given as the product of the separate likelihoods of the data (we assume they are statistically independent) as follows:
$$
L_{\rm tot}=L_{\rm SnIa} \times L_{\rm BAO} \times L_{\rm H(z)} \times L_{\rm CMB},
$$
which is related to the total $\chi^2$ via $\chi^{2}_{\rm tot}=-2\log{L_{\rm tot}}$ or
\be
\chi^{2}_{\rm tot}=\chi^{2}_{\rm SnIa}+\chi^{2}_{\rm BAO}+\chi^{2}_{\rm H(z)}+
\chi^{2}_{\rm CMB}.\label{eq:chi2eq}
\ee
In order to study the statistical significance of our constraints we make use of the well known Akaike Information Criterion (AIC)~\cite{Akaike1974}. Assuming Gaussian errors, the AIC parameter is given by
\begin{eqnarray}
{\rm AIC} = -2 \ln {\cal L}_{\rm max}+2k_p+\frac{2k_p(k_p+1)}{N_{\rm dat}-k_p-1} \label{eq:AIC}\;,
\end{eqnarray}
where $k_p$ and $N_{\rm dat}$ are the number of free parameters and the total number of data points respectively. In this analysis we have 1048 data points from the Pantheon set, 4 from the CMB shift parameters, 10 from the BAO measurements and finally 36 $H(z)$ points, for a total of $N_{\rm dat}=1098$.

The AIC can be interpreted similarly to the $\chi^2$, i.e. a smaller relative value signifies a better fit to the data. To apply the AIC to model selection we then take the pairwise difference between models $\Delta {\rm  AIC}={\rm AIC}_{\rm model}-{\rm AIC}_{\rm min}$. This is usually interpreted via the Jeffreys' scale as follows: when $4<\Delta {\rm AIC} <7$ this indicates positive evidence against the model with higher value of ${\rm AIC}_{\rm model}$, while in the case when $\Delta {\rm AIC} \ge 10$ it can be interpreted as strong evidence. On the other hand, when $\Delta {\rm AIC} \le 2$, then this means that the two models are statistically equivalent. However, Ref.~\cite{Nesseris:2012cq} has shown that the Jeffreys' scale can lead to misleading conclusions, thus it should always be interpreted carefully.

Then, our total $\chi^2$ is given by Eq.~(\ref{eq:chi2eq}) and the parameter vectors (assuming a spatially flat Universe) are given by: $p_{\Lambda \textrm{CDM}}=\left(h, N_{\rm eff}, \Omega_{b,0}, \Omega_{c,0}\right)$ for the $\Lambda$CDM;
and  $p_{\rm PBH}=\left(h, N_{\rm eff}, \Omega_{b,0}, \Omega_{c,0}, \alpha, f_{\rm PBH} \right)$ for the PBH model.
For the last two parameter, we will actually use the parameter $\log_{10}\alpha$ and $\log_{10}f_{\rm PBH}$ in order to sample the parameter space much better, given that we expect to have small values of both parameters.

Using the aforementioned cosmological data and methodology, we can obtain the best-fit parameters and their uncertainties via the MCMC method based on a Metropolis-Hastings algorithm. The codes used in the analysis were written independently by two of the authors, in both Mathematica and Python 3.0.\footnote{The MCMC code for Mathematica used in the analysis is freely available at \url{http://members.ift.uam-csic.es/savvas.nesseris/}.} The priors we assumed for the parameters are given by $h \in[0.4, 1]$, $N_{\textrm{eff}} \in [1,5]$, $\Omega_{b,0} \in[0.001, 0.1]$, $\Omega_{c,0} \in[0.1, 1]$, $\log_{10}\alpha \in[-20, 0]$
$\log_{10}f_{\rm PBH} \in[-20, 0]$. We estimate $\sim10^5$ MCMC points for each of the two models.

As a further step we decide to fix both $\log_{10}\alpha = -4$ to which corresponds a primordial black hole mass of $M = 9.79\cdot 10^{15}$g,  and $\log_{10}f_{\rm PBH}= 0$ in order to have the largest contribution in the PBH scenario and study the contribution to the overall  dynamics of the Universe.

\subsection{Results}
In Fig.~\ref{fig:MCMCPBH-full} we show  the 68.3$\%$, 95.4$\%$ confidence contours, long with the 1D marginalized likelihoods for various parameter combinations, for all the six parameters entering in the PBH scenario, i.e.  $p_{\rm PBH}=\left(h, N_{\rm eff}, \Omega_{b,0}, \Omega_{c,0}, \log_{10}\alpha, \log_{10}f_{\rm PBH} \right)$ and in Tab.~\ref{tab:bestfits} we report their best fit. We considered two separate cases, first using only the CMB data and then using all the data together.
The data used are clearly insensitive to the PBH parameters, as the full marginalized errors span over the whole range of the values allowed in the analysis.
The reason is twofold: 1) the CMB shift parameters have been evaluated using only the TT modes of the CMB which are particularly insensitive to the PBH physics as also evidenced by~\cite{Stocker:2018avm}; 2) the effective dark energy equation of state, Eq.~\eqref{eq:w_eff_PBH}, manifests a very mild phantom behavior at late time, making it practically undistinguishable from $w=-1$.

In Fig.~\ref{fig:MCMCPBH} we show the 68.3$\%$ and 95.4$\%$ confidence contours for the $\Lambda$CDM and the PBH models, respectively, along with the 1D marginalized likelihoods for various parameter combinations. The PBH results have been obtained by fixing
$\log_{10}\alpha = -4$ and $\log_{10}f_{\rm PBH}= 0$. In Tab.~\ref{tab:bestfits} are reported the corresponding best fit. Also in this analysis we considered two different cases, CMB data alone and all the data together.
In this analysis we wanted to have the largest contribution possible from the PBH but still within the allowed regions reported in~\cite{Stocker:2018avm}. In this case, we do not see any appreciable difference on the best fit of the parameters. Both models, i.e. $\Lambda$CDM and PBH give very similar results, implying that the contribution of energy budget from PBH does not affect the expansion of the Universe. We want to highlight that we did not expect any change on the best fit of the parameters from CMB data alone, because the initial conditions of our dynamical equations were set to be exactly $\Lambda$CDM.

In Table~\ref{tab:chi2AIC} we show the values for the $\chi^2$ and AIC parameters for the $\Lambda$CDM and the PBH models respectively. As mentioned, we  considered two separate cases: CMB data alone, and all the data together. In the first case, by inspecting Tables \ref{tab:bestfits} and \ref{tab:chi2AIC}, we find that as the difference in the AIC parameters is roughly 0 and $\lesssim 1.3$ for the PBH and full PBH models with respect to the \lcdm model, then they all are in good agreement with each other. When we use all the data, we find that the statistical difference rises to $\sim2$ and $\sim7$ for the PBH and full PBH models, thus placing some strain on the full PBH case with respect to the \lcdm model.

\subsection{Speculative venues}
Let us now  ask the question of what would be the effect of an ultra-light PBH fraction on the expansion history. For masses below around $10^{14}$ g such a population would not serve as DM since it would have completely evaporated by now. Nevertheless, in this case the radiation injection could be enough to produce a change in $N_{\rm eff}$ and consequently affect $H_0$. For a fraction of $f_{\rm PBH}=10^{-7} $, a mass of $M_{\rm PBH}=10^9$ g could achieve this\footnote{For these ultra-light black holes the value of the constant $\C$ appearing in Eq.~\eqref{eq:eq:mprime} is larger by a factor of $\sim15$~\cite{Carr:2009jm}. This is because the lighter the BH, the higher its Hawking temperature and hence the heavier the particles emitted. In fact, for this mass, all Standard Model particles contribute to the radiation yielding $F(M)\sim 67.54 \times 10^{-4}$.}; they would start radiating around a temperature of $T=T_H\sim 10\;$TeV and would emit the bulk of their mass around neutrino decoupling pumping relativistic particles into the plasma. In fact, we can have an increase of
\be \label{DNeff}
\Delta N_{\rm eff}=\frac{ \rho_{\rm x} }{ \rho_{\nu}  }\simeq \frac87 \left(\frac{11 }4 \right)^{\frac43} \frac{ \rho_{\rm x} }{ \rho_{\gamma}  } = 0.35,
\ee
which, as a result, raises the $H_0$ value. Such ultra-light black holes are compatible with mass constraints~\cite{Carr:2009jm} if no physics beyond the Standard Model is assumed at this scale, since they have no impact on the CMB anisotropies~\cite{Stocker:2018avm}, neither do they affect CMB spectral distortions~\cite{Poulin:2016anj}.

\begin{figure}[b]
\includegraphics[width = 0.35\textwidth]{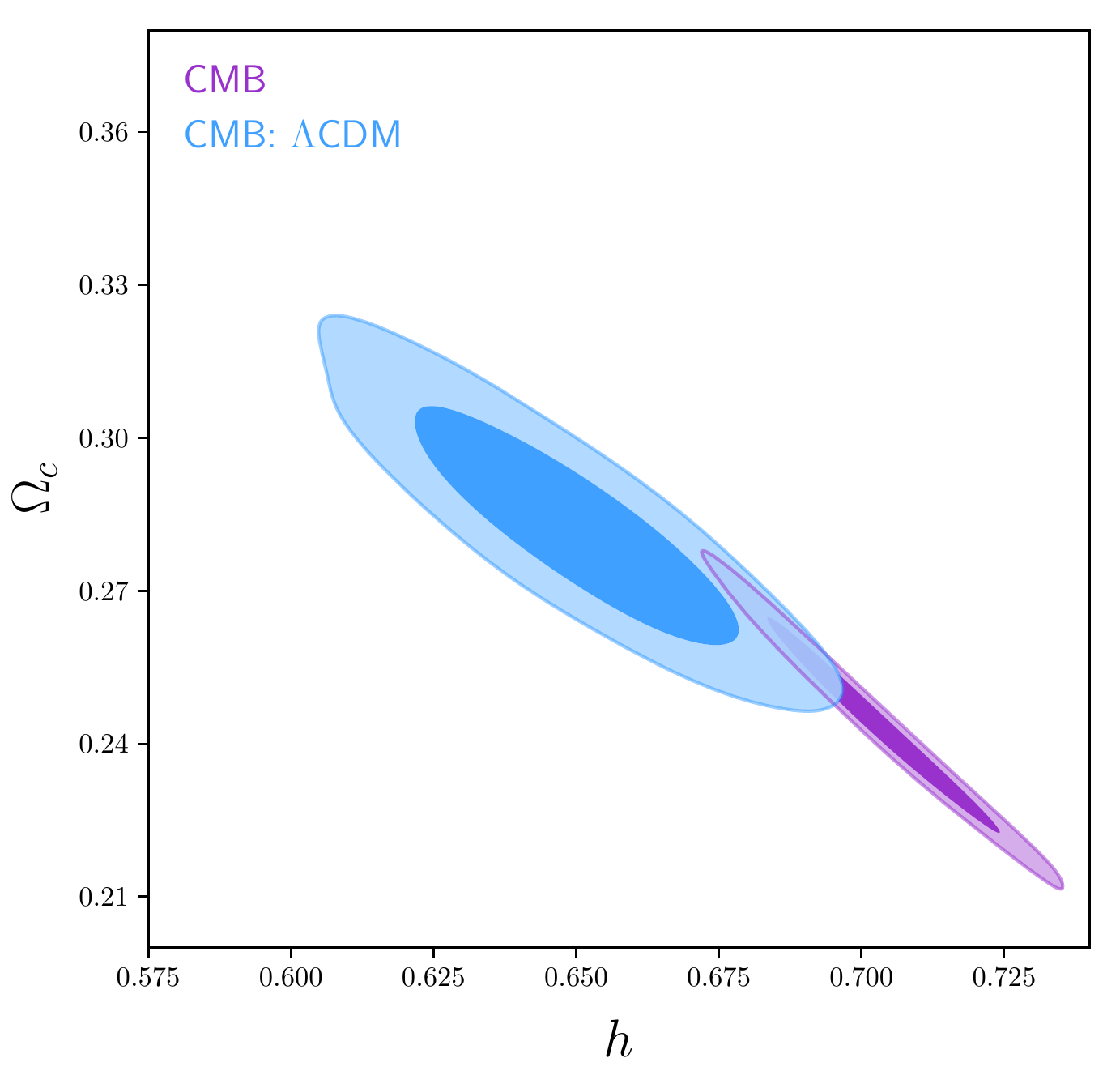}
\caption{The 68.3$\%$ and 95.4$\%$  confidence contours for $h-\Omega_{c,0}$ only for ultra-light PBH and the $\Lambda$CDM model for comparison.
\label{fig:MCMCPBH-spec}}
\end{figure}

In Fig.~\ref{fig:MCMCPBH-spec}, we show the confidence regions for $h-\Omega_{c,0}$ using CMB data only. For such ultra-light PBH, there is a substantial shift on $H_0$ up to $70.49$, reducing the tension with local measurements. The best fit parameters with their 1$\sigma$ errors are:
\bea
h &=& 0.7049 \pm 0.0134\nonumber\,, \\
\Omega_{b,0} &=&  0.0448 \pm 0.0015\nonumber\,, \\
\Omega_{c,0} &=&  0.2415 \pm 0.0139\nonumber\,.
\eea
The minimum $\chi^2$ is $0.0170$ and the AIC criterium gives $-9.8546$. Compared with Tab.~\ref{tab:chi2AIC} we find a difference on $\Delta $AIC of about $4$ which indicates a positive evidence in favor of $\Lambda$CDM. Interestingly,  even though there is a shift on the values of the parameters, their products give a result very close to the Planck best fit within the $1\sigma$ errors,~\cite{Aghanim:2018eyx}:  $\Omega_{c,0} h^2 = 0.1199$ and $\Omega_{b,0} h^2 = 0.02226$.

Let us note, however, that a more careful analysis is needed in order to draw more precise conclusions. For example, the last equality in Eq.~\eqref{DNeff} holds for redshifts beyond the neutrino decoupling, which is the lifetime of a $10^9\;$g black hole. That would be an excellent approximation for a $10^{10}\;$g population decaying well beyond this time, such PBHs though are severely constrained since they start affecting the light elements abundance~\cite{Carr:2009jm}. A more accurate approach would be to compute the change in the number of relativistic species using the methods of~\cite{Stocker:2018avm} for the relevant mass range, which we leave for future work.

It would be also interesting to further investigate the mechanisms that produce such ultra-light PBHs on top of the DM candidate population.
For instance, assuming that their production mechanism is inflationary in origin, the mass of the BH depends on the e-folds that a mode spends outside the horizon and the energy scale of inflation~\cite{Garcia-Bellido:2017mdw}. In particular, $10^9\;$g PBHs can be produced via both single field inflation~\cite{Garcia-Bellido:2017mdw} and critical Higgs inflation~\cite{Ezquiaga:2017fvi} scenarios. In both cases, an inflection point has to be present at the potential and traversed by the field roughly $20-40$ e-folds before the end of inflation. The superhorizon modes responsible for such PBHs are the ones experiencing exponential growth during the last $9$ e-folds of the plateau. Another possibility of an ultra-light PBH abundance is when they are produced via resonant instabilities due to reheating~\cite{Martin:2019nuw} or features~\cite{Cai:2018tuh}, from collapsing modes that spend $7-8$ e-folds outside the horizon.

\section{Conclusions \label{conclusions}}
Primordial black holes present a paradigm shift in our understanding of the nature of dark matter. In this paper, we presented a novel scenario where due the quantum effects of Hawking evaporation of PBH, the late time dynamics of the expansion of the Universe is affected via this new radiation component. The method consists in coupling the BH density to the new radiation component, assuring that the energy loss by the black holes is transferred to radiation.

The idea was to understand and test how the cosmological dynamics is affected by the presence of light PBH, compatible with DM, focusing on the $H_0$ problem. Our analysis showed that, even if there is a constant pumping of energy into the radiation sector, it is not able to reconcile the long/short $H_0$ discrepancy at 2$\sigma$, if black holes account for the DM. TT modes from CMB data still prefer a value of $N_{\rm eff}$ close to the $\Lambda$CDM value, depriving the effective number of relativistic species to increase and hence to modify the $H_0$ value.

Furthermore, we found that for masses in the allowed asteroid range, where PBHs can serve as DM, the decay mechanism can be formulated as an effective dark energy fluid, and hence it can be interpreted as a late time effect. The physics behind is that the emitted particles do not behave as radiation even though they are relativistic, since $\omx$ does not scale exactly as $a^{-4}$ due to Eq.~\eqref{eq:eq:mprime}. By solving the system of coupled fluids numerically, it can be shown that at late times the radiation component shows a phantom behavior, being too mild, though, to differ appreciably from the cosmological constant.

Nevertheless, this formulation allowed us to consider the effect of a fraction of matter ($f_{\rm PBH}\sim10^{-7}$) in tiny PBHs ($M_{\rm PBH}\sim10^9\;$g). Such black holes cannot explain the DM abundance, however, their complete decay around neutrino decoupling can sufficiently alter the number of relativistic species, thus, raising the CMB induced current Hubble rate to $H_0\simeq70.5$.

Finally, using our approach and comparing with the latest cosmological data, including the SnIa, BAO, CMB and the H(z) data, we placed observational constraints on the PBH model. We found that the PBH model is statistically consistent with \lcdm according to the AIC statistical tool.

\section*{Acknowledgements}
The authors would like to thank Malcolm Fairbairn, Martiros Khurshudyan, Gonzalo Palma and Davide Racco for fruitful discussions and comments. S.N. acknowledges support from the Research Project FPA2015-68048-03-3P [MINECO-FEDER], PGC2018-094773-B-C32 and the Centro de Excelencia Severo Ochoa Program SEV-2016-0597, use of the Hydra cluster at the IFT and support from the Ram\'{o}n y Cajal program through Grant No. RYC-2014-15843. SS is supported by the CUniverse research promotion project (CUAASC) at Chulalongkorn University.\\~~\\

\appendix
\section{Differential equations}
In this section we report the details of the analysis performed. The system of differential equations used is
\bea
\tilde{\Omega}_{\rm PBH}'(a) &=& -\frac{\alpha}{a\,H}\frac{(f_{\rm PBH}\Omega_{c,0})^3}{\tilde{\Omega}_{\rm PBH}(a)^2}\label{eq:oPBH_tilde-app}\,,\nonumber\\
\tilde{\Omega}_{\rm x}'(a) &=& \frac{\alpha}{a^{1-3w_{\rm x}}\,H}\frac{(f_{\rm PBH}\Omega_{c,0})^3}{\tilde{\Omega}_{\rm PBH}(a)^2}\label{eq:ox_tilde-app}\,,\nonumber
\eea
where we have set $\alpha = \C/M^3_{\rm in}$. The Hubble parameter for solving the above system should have the expression of Eq.~\eqref{eq:Hubble-op}. However, the former contains the cosmological term $\Omega_{\Lambda,0}$ which has to be treated as dependent parameter in order to keep the unitary of the Hubble parameter at $z=0$, i.e. $H(z)/H_0=1$. Consequently, we need to impose
\bea
\Omega_{\Lambda_0} &=& 1-\Omega_{b_0}-(1-f_{\rm PBH})\Omega_{c_0}-\Omega_{r_0}  \nonumber \\
&-&\op(a=1)+\Omega_{\rm x}(a=1)\,.\nonumber
\eea
Clearly the above expression cannot be used to solve the differential equations for $\tilde{\Omega}_{\rm PBH}$ and $\tilde{\Omega}_{\rm x}$, as it would require a value of the energy densities at the $a=1$.
To overcome this difficulty, we adopt the following strategy: we defined an internal Hubble parameter of the form of Eq.~\eqref{eq:Hubble-op} where $\Omega_{\Lambda_0} = 1-\Omega_{b_0}-\Omega_{c_0}-\Omega_{r_0}$.
This approximation is sufficient for our analysis especially for the range of allowed PBH masses; in fact for a mass of $10^{16}$g it would require $\approx 10^{15}$ years for the black hole to completely evaporate.

Furthermore, the analysis has been also tested with a brute method, i.e. forcing the system to give the appropriate $\Omega_{\Lambda_0}$ in order to keep the normalized Hubble parameter equal to $1$.
We did not see any change on the final results and we adopted the strategy mentioned above for the MCMC analysis.

\bibliographystyle{utphys}
\bibliography{biblio_pbh}

\end{document}